\documentclass[sigconf]{acmart}
\usepackage{subcaption}
\usepackage{tabularx}

\usepackage{enumitem}
\newenvironment{tight_itemize}{
\begin{itemize}[leftmargin=20pt]
  \setlength{\topsep}{0pt}
  \setlength{\itemsep}{0pt}
  \setlength{\parskip}{0pt}
  \setlength{\parsep}{0pt}
}{\end{itemize}}

\usepackage[autostyle]{csquotes}
\copyrightyear{2026}
\acmYear{2026}
\setcopyright{cc}
\setcctype{by}
\acmConference[ICMI Companion '26]{Companion of the INTERNATIONAL CONFERENCE ON MULTIMODAL INTERACTION}{October 05--09, 2026}{Napoli, Italy}
\acmBooktitle{Companion of the INTERNATIONAL CONFERENCE ON MULTIMODAL INTERACTION (ICMI Companion '26), October 05--09, 2026, Napoli, Italy}
\acmDOI{10.1145/3776591.3838179}
\acmISBN{979-8-4007-2319-3/2026/10}

\begin{document}

% --- Title and Authors ---
\title[Embodied Empathy for Self-Attachment Psychotherapy with Self‑Initiated Humour]{Embodied Empathy: A Multimodal AR and LLM-Powered System for Self-Attachment Psychotherapy with Self‑Initiated Humour}

\author{Xinyan Ye}
\affiliation{%
  \institution{Imperial College London}
  \city{London}
  \country{UK}
}
\email{xinyan.ye19@imperial.ac.uk}

\author{Gwyneth Phang}
\authornote{This work was done while this author was a student at Imperial College London.}
\affiliation{%
  \institution{Imperial College London}
  \city{London}
  \country{UK}
}
\email{gwynethphang@gmail.com}

\author{Anandha Gopalan}
\affiliation{%
  \institution{National University of Singapore}
  \country{Singapore}
}

\email{axgopala@nus.edu.sg}
\author{Abbas Edalat}
\affiliation{%
  \institution{Imperial College London}
  \city{London}
  \country{UK}
}
\email{a.edalat@imperial.ac.uk}

% --- Short Authors ---
% This ensures the header reads "Phang et al." instead of a long list
\renewcommand{\shortauthors}{Ye et al.}

% --- Abstract (Must be before maketitle) ---
\begin{abstract}
The growing global demand for mental health support increasingly exceeds the supply of qualified practitioners, creating an urgent need for scalable digital interventions that can deliver meaningful emotional connection. In response, we present a novel multimodal application that operationalises the Self‑Initiated Humour Protocol (SIHP) within a Self‑Attachment Technique (SAT) framework. Our mobile application integrates customisable 3D childhood avatars, augmented reality, and an LLM‑driven virtual therapist capable of automated emotion mirroring. An eight‑day user study (N=16) indicates the system’s feasibility and improvements in self-reported mood. Results show that personalised avatars and text‑to‑speech output strengthen emotional bonding and perceived empathy. Although emotion mirroring boosts engagement, its effectiveness depends heavily on classification accuracy and animation intensity. Moreover, findings indicate a shift in user expectations—from reactive chatbots to proactive conversational facilitators. We conclude with design implications for leveraging AI and AR to cultivate embodied empathy in digital mental health tools.
\end{abstract}

% --- CCS Concepts (Mandatory for ACM) ---
% You must generate these at http://dl.acm.org/ccs.cfm


% --- Keywords ---
% TODO: add to this

% \citestyle{acmauthoryear}

% \begin{teaserfigure}
%   \centering
%   \begin{subfigure}[b]{0.49\textwidth}
%     \centering
%     \includegraphics[height=5.7cm]{figures/teaser_vtr_resize.png}
%     \caption{}
%     \label{fig:teaser_vtr}
%   \end{subfigure}
%   \hfill
%   \begin{subfigure}[b]{0.49\textwidth}
%     \centering
%     \includegraphics[height=5.7cm]{figures/teaser_ar_resize.png}
%     \caption{}
%     \label{fig:teaser_ar}
%   \end{subfigure}
%   \caption{The mobile iOS application enables users to practice self-attachment psychotherapy through (a) multimodal interaction with a childhood self avatar and a therapist avatar. (b) By projecting the childhood avatar into an AR environment, the system facilitates spatial co-presence, fostering a foundational emotional and empathic bond with one's childhood self. {\small \textit{[Teaser figures generated by Google Gemini (text-to-image) \cite{team2023gemini}]}}}
%   \label{fig:teaser}
% \end{teaserfigure}

\maketitle
% \begin{figure}[H]
%     \centering
%     % 
%     \begin{subfigure}[b]{0.495\textwidth}
%         \centering
%         \includegraphics[width=\textwidth]{figures/teaser_vtr_resize.png}
%         \caption{}
%         \label{fig:teaser_vtr}
%     \end{subfigure}
%     \hfill % 
%     \begin{subfigure}[b]{0.495\textwidth}
%         \centering
%         \includegraphics[width=\textwidth]{figures/teaser_ar_resize.png}
%         \caption{}
%         \label{fig:teaser_ar}
%     \end{subfigure}
    
%     \caption{The mobile iOS application enables users to practice self-attachment psychotherapy through (a) multimodal interaction with a childhood self avatar and a therapist avatar. (b) By projecting the childhood avatar into an AR environment, the system facilitates spatial co-presence, fostering a foundational emotional and empathic bond with one's childhood self. {\small \textit{[generated by Google Gemini (text-to-image)]}}}
%     \label{fig:teaser}
% \end{figure}

\begin{figure*}[t]
    \centering
    \begin{subfigure}[b]{0.45\textwidth}
        \centering
        \includegraphics[height=6.5cm]{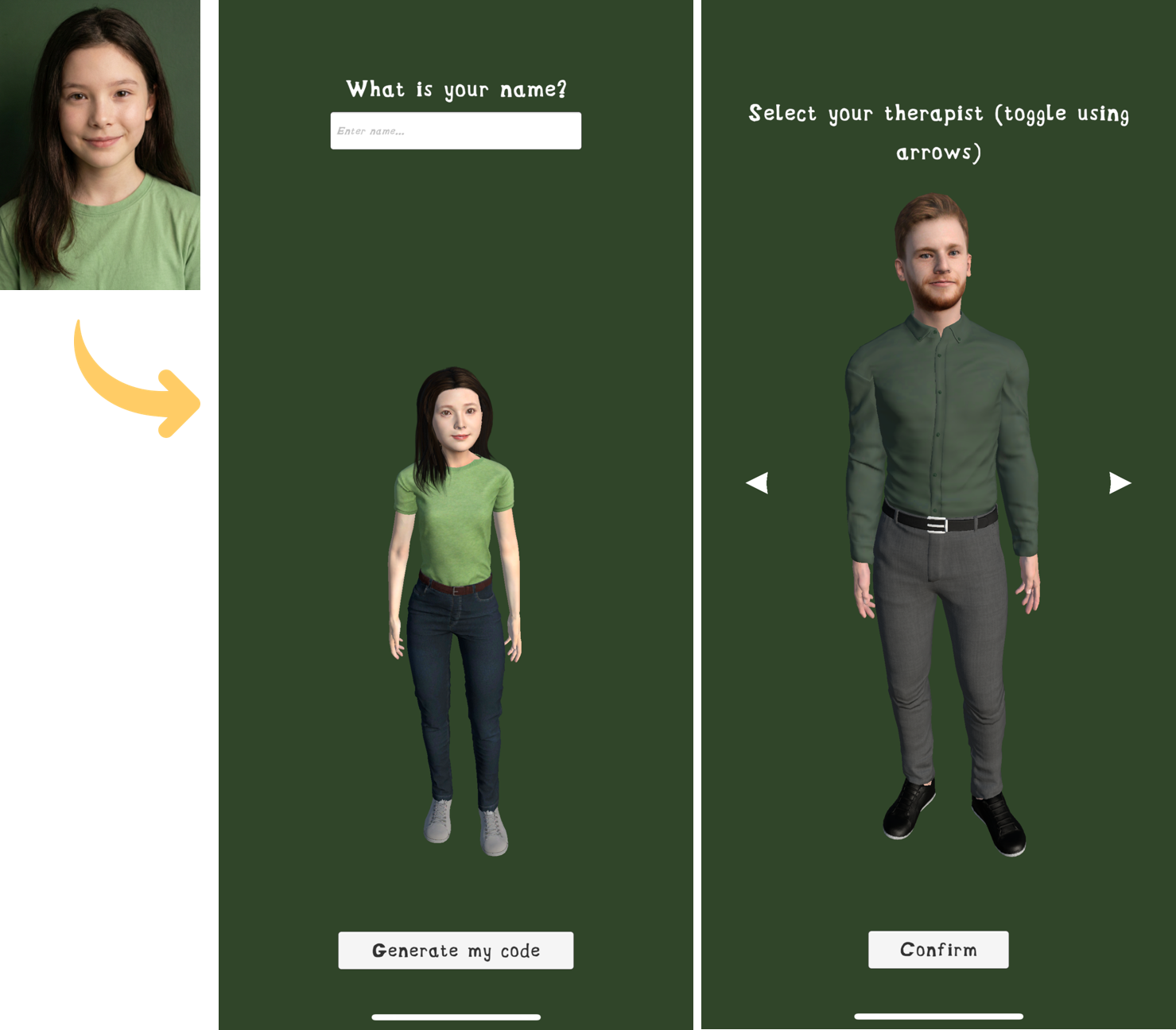}
        \caption{Child Avatar Creation and Therapist Avatar Selection}
        \label{fig:avatarcreate}
    \end{subfigure}
    \hfill
    \begin{subfigure}[b]{0.19\textwidth}
        \centering
        \includegraphics[height=6.5cm]{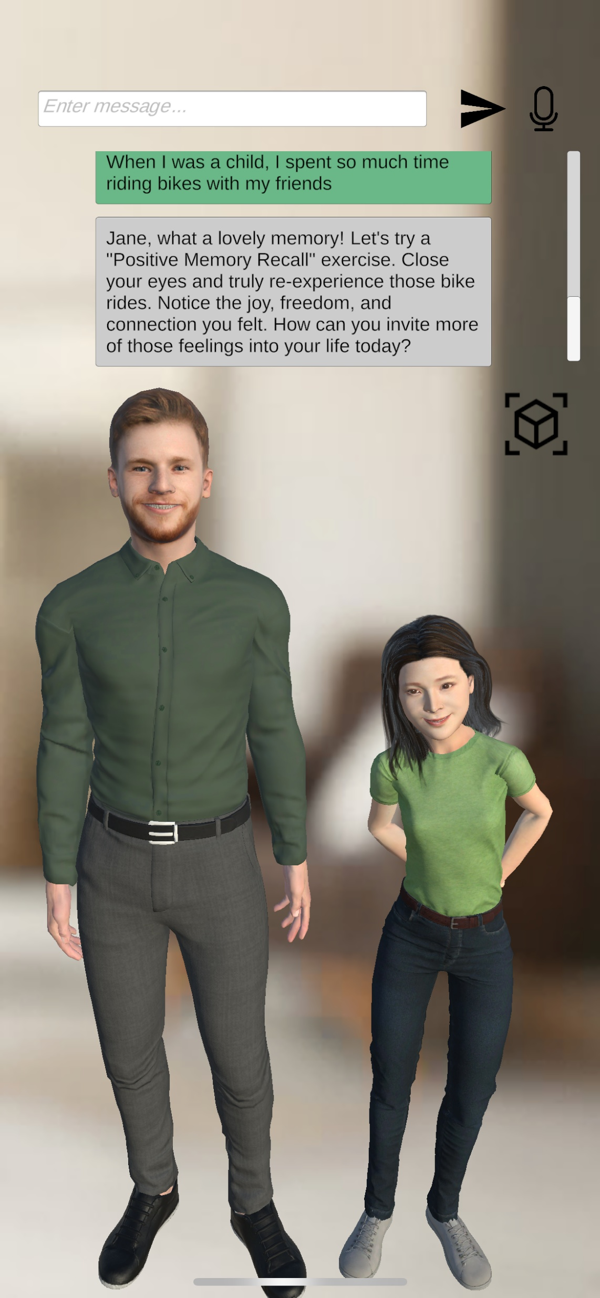}
        \caption{Virtual Therapy Room}
        \label{fig:VTR}
    \end{subfigure}
    \hfill
    \begin{subfigure}[b]{0.22\textwidth}
        \centering
        \includegraphics[height=6.5cm]{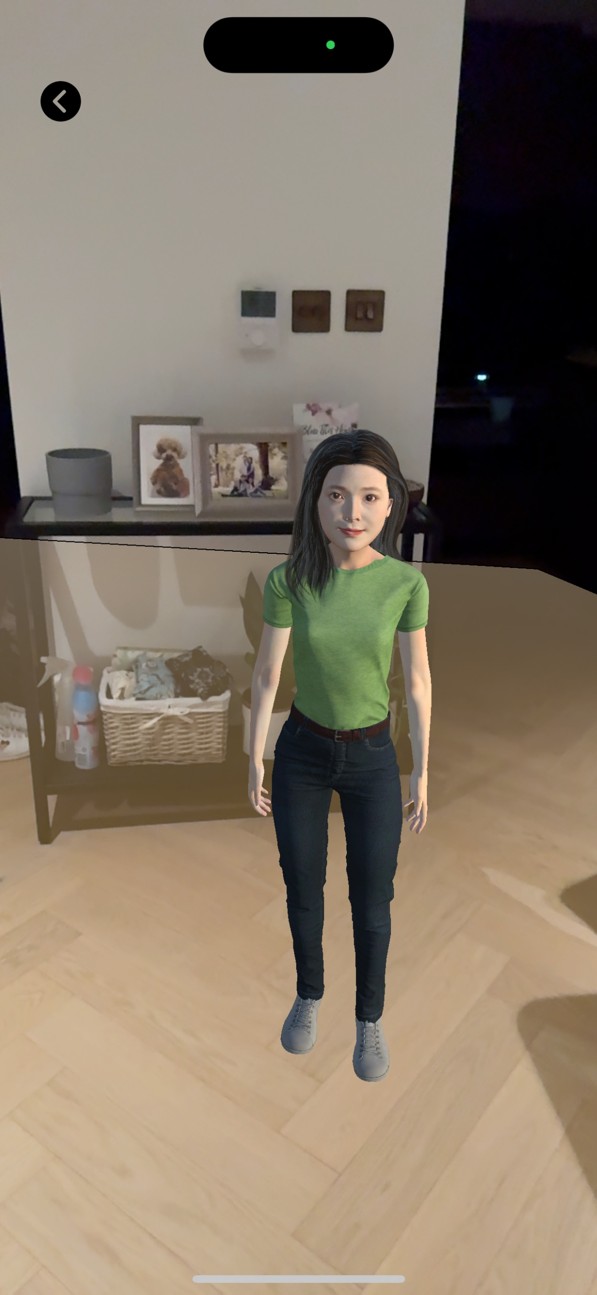}
        \caption{AR Mode}
        \label{fig:armode}
    \end{subfigure}
    \caption{Key features and interfaces of the mobile application: (a) Avatar Creation and Selection, allowing users to generate a personalised childhood avatar via photo-upload and select a preferred therapist avatar; (b) Virtual Therapy Room, showing the multimodal chatbot interface and real-time emotional mirroring across both avatars; and (c) AR Mode, enabling spatial co-presence by projecting the childhood avatar into the user's physical environment.}
    \label{fig:app_feats}
\end{figure*}

%%%%%%%%%%%%%%%%%%%%%%%%%%%%%%%%%%%%
\section{Introduction}

One in eight people in the world live with mental disorders, defined by a significant impact on cognitive abilities, emotional regulation, or behaviour \cite{who2022mental}. However, this number only accounts for clinically diagnosed mental disorders. Mental health extends beyond the absence of mental disorders; it exists on a continuum and is experienced to varying degrees of difficulty and distress. Despite the large scale and impact of mental health, most people do not have access to effective care largely due to socio-economic factors, resulting in a persistent \enquote{mental health treatment gap} \cite{who2022mental_health_strengthening}.

To bridge this gap, digital mental health interventions, particularly online therapy, have emerged as a promising approach. These platforms reduce costs, eliminate travel barriers, and alleviate the discomfort or stigma associated with face-to-face help-seeking, thereby increasing engagement \cite{van2024clients}. Consequently, chatbot-assisted interventions have gained rapid popularity due to their scalability, low barriers to entry, and continuous availability \cite{lim2022chatbot}. Early mobile mental health applications, such as Woebot \cite{woebot2026official} and Wysa~\cite{wysa2026official}, relied primarily on pre-defined decision trees and text-based Cognitive Behavioural Therapy (CBT).  Prior analyses suggest that such chatbots can reduce depressive symptoms among adults with anxiety and depression, with effects comparable to traditional psychotherapy and outperforming non-directive treatments \cite{fitzpatrick2017delivering}. However, clinicians have expressed concerns that text-only interactions struggle to provide care that feels genuine, personalised, and empathic \cite{moylan2025expert}.

Recently, the digital mental health field has reached a pivotal inflection point. In June 2025, Woebot, which once served over 1.5 million users, announced the shutdown of its consumer service~\cite{woebot2026shutting}. Despite  multiple clinical trials proving its efficacy \cite{fitzpatrick2017delivering,mariano2021therapeutic}, its sunset highlighted a core pain point in digital therapeutics: rule-based, text-driven systems struggle to maintain user engagement over time and face immense regulatory and business pressures due to the rise of generative AI, particularly large language models (LLMs)~\cite{maheu2025aipsychotherapy}. This event signals a transition in \enquote{AI psychotherapy} from rigid, scripted dialogues toward flexible, generative interactions that prioritise emotional connection and engagement.

From the perspective of interaction science, while LLMs generate high-quality, context-aware text that appears empathic, there are limitations of pure text. Non-verbal cues are a critical component of therapeutic alliance \cite{park2018correlation} and are positively correlated with successful clinical outcomes \cite{kraft2017empathic}. Humans rely heavily on facial expressions, posture, vocal tone, and spatial presence to establish a sense of safety and support and text-based emotional support often falls short  in providing users with deep emotional regulation. While applications like Replika \cite{replika2026official} or Character.AI \cite{characterai2026} use 3D avatars to enhance immersion and social presence, they are designed for general social interaction rather than targeted psychological intervention. Concurrently, research has explored immersive media (VR/AR) and embodied agents for healthcare. While VR has shown success in treating anxiety, depression, substance abuse and eating disorders \cite{yeung2021virtual}, it faces a significant barrier to entry due to cost, accessibility to equipment, health concerns (cybersickness), misconceptions, and limited accessibility for users with motor and sensory impairments \cite{felnhofer2025barriers,firdaus2024factors,ahmed2025designing,hartfill2025embracing}. In contrast, AR can be deployed on standard smartphones. By overlaying virtual elements onto the physical environment, AR offers \enquote{spatial co-presence} without isolating the user from reality, supporting more natural embodied interactions \cite{963459, baus2014moving}. However, integrating this spatiality with specific psychotherapeutic theories, particularly those reliant on emotional bonds and visual representation, remains under-explored.

In this context, interventions based on the Self‑Attachment Technique (SAT)~\cite{edalat2017self,edalat2015introduction,edalat2024affectional} offer a promising foundation for embodied digital interaction. SAT, extended through the Self‑Initiated Humour Protocol (SIHP)~\cite{edalat2022sihp,RefWorks:edalat2025self-initiated}, aims to foster an affectionate bond between individuals and their childhood selves while guiding them to practice non‑hostile laughter in varied life circumstances~\cite{RefWorks:edalat2025self-initiated}. Although humour‑based interventions have been associated with positive mental‑health outcomes, their mechanisms are often treated as a \enquote{black box}, with limited procedural detail~\cite{kafle2023beyond_laughter}. SIHP addresses this gap by explicating each humour rule through established theories of laughter. Prior SIHP trials used a personalised VR avatar using Google Cardboard, paired with a rule‑based, emotionally intelligent chatbot accessed via a website~\cite{RefWorks:edalat2025self-initiated}. While results were encouraging, participants noted critical limitations: lack of open‑ended, flexible dialogue and fragmented experience between VR avatar and web‑based chatbot. This separation undermined therapeutic cohesion and restricted opportunities for deeper, empathic engagement.

To address this technical gap, we built a mobile application that offers an integrated, immersive platform for SAT/SIHP practice. This application enables users to create a customisable \enquote{childhood-self} avatar, viewable in a virtual therapy room or placed within an AR environment, and replaces the previous rule-based chatbot with a multimodal LLM-driven virtual therapist. By unifying LLM-based dialogue, customisable avatars, voice interaction (Text-to-Speech and Speech-to-Text), and automatic emotion recognition and mirroring, the platform aims to provide a more engaging, supportive, and emotionally containing environment for SIHP.

Our research investigates how different interaction modalities within the application
influence user engagement, perceived empathy, and affective responses and shape the user experience of SAT with SIHP. We hypothesise that:
(1) enabling personalisation of a childhood avatar and AR interaction will strengthen emotional bonding with the user's childhood self;
(2) emotion mirroring through the avatars will enhance emotional connection and foster a sense of embodied empathy; and
(3) a voice-enabled therapist (via speech and audio feedback) will support more expressive practice during SAT/SIHP exercises as compared to text-only interaction.

%Specifically, we examine how (1) personalised childhood avatars and AR interaction, (2) emotion mirroring in both the child and therapist avatars, and (3) a voice-enabled virtual therapist influence user engagement, perceived empathy, and affective responses.

We conducted an eight-day remote user study (N = 16) with non-clinical adult users to evaluate the feasibility, acceptability, and experiential qualities of this multimodal SAT/SIHP delivery. Our results reveal several key insights. (1) Personalised child avatars emerge as central to emotional bonding, while AR amplifies that connection in short, intentional episodes. (2) Emotion mirroring is powerful but misclassifications or overly intense animations can disrupt the therapeutic experience. (3) Voice output (Text-to-Speech, TTS) substantially enhances perceived empathy and containment, whereas voice input (Speech-to-Text, STT) offers marginal benefits. (4) Participants strongly valued conversational proactivity, which shows that the LLM-driven therapist persona should act as a facilitator, actively guiding the process instead of merely offering passive responses. To summarise, our contributions are:

\begin{enumerate}
    \item \textbf{A novel multimodal SAT/SIHP platform:} We present the design and implementation of a mobile application that unifies SAT/SIHP delivery into a single experience, integrating customisable child avatars that can be rendered in both a virtual therapy room and AR, an LLM-driven therapist avatar with emotion mirroring, and voice interaction (STT \& TTS) to deliver SAT/SIHP in an embodied, immersive format.

    \item \textbf{Feasibility of hybrid avatar–chatbot–AR therapy:} We provide evidence that the application is both feasible and accepted by non-clinical users. Quantitative and qualitative data indicate a positive impact on user self-reported mood and high levels of therapeutic engagement, with 87.5\% of participants endorsing the application as a viable tool for emotional self-regulation.

    \item \textbf{Empirical insights into modality-specific trade-offs:} We identify the distinct roles of multimodal features: (i) personalised childhood avatars serve as the primary anchor for self-attachment, (ii) AR provides high-impact spatial presence, and (iii) TTS is a more potent driver of perceived empathy than STT, providing clear guidance on where to prioritise design efforts in future digital therapeutics.

    \item \textbf{Shifting User Expectations Toward AI Proactivity:} We uncover a critical evolution in users' relationship  with AI psychotherapy. Users transition from expecting reactive chatbot responses to seeking proactive conversational facilitators that actively guide the therapeutic process.

\end{enumerate}

%%%%%%%%%%%%%%%%%%%%%%%%%%%%%%%%%%%%
\section{Background}

\subsection{Self-attachment and Self-Initiated Humour Protocol (SAT/SIHP)}
\label{sec:SIHP}

SAT/SIHP is a two-step self-administered protocol that has undergone a successful pilot study \cite{RefWorks:edalat2025self-initiated}. The first stage comprises Self-Attachment Technique (SAT) exercises rooted in attachment theory. As insecure childhood attachments have been linked to affective disorders in adulthood \cite{mikulincer2012attachment}, SAT exercises aim to establish neural patterns associated with secure attachment to foster social and emotional maturity \cite{edalat2017self,edalat2022sihp}. This is achieved by interacting with one's childhood self while simultaneously assuming caregiving and care-seeking roles, building a compassionate bond between the adult and child self \cite{edalat2024affectional}. These exercises were delivered via a personalised mobile VR application in which users bond with a childhood avatar generated from submitted images, practising individually for at least 15 minutes twice a day over the 8-week intervention \cite{edalat2024affectional}.

The second stage consists of SIHP exercises, which build on SAT. Whereas SAT focuses on caring for one's childhood self, SIHP expands on SAT's basic laughter exercises, enabling users to find humour across all life circumstances~\cite{RefWorks:edalat2025self-initiated}. Drawing on the primary theories of humour (Superiority, Incongruity, Play, and Evolutionary), SIHP exercises are practised independently with the guidance of an emotionally intelligent chatbot \cite{edalat2022sihp,RefWorks:edalat2025self-initiated}. Unlike traditional interventions that rely on jokes or intentional laughter, SIHP's novelty lies in its capacity to fundamentally reshape the individual's mindset and outlook on life. This humour-driven expansion of the SAT framework has been shown to significantly enhance overall well-being, self-compassion, and self-enhancing humour \cite{RefWorks:edalat2025self-initiated}.

% Research has demonstrated that SIHP can significantly improve well-being, self-compassion and  self-enhancing humour. 

% Previous SIHP trials were carried out through a personalised VR avatar and  an emotionally intelligent, rule-based chatbot accessed via a website \cite{RefWorks:edalat2025self-initiated}. Despite generally positive outcomes and acceptance of the chatbot-VR delivery, a major limitation is the fragmented user experience: participants had to switch between a web-based chatbot and a Google Cardboard VR experience. This discontinuity disrupted affective flow and therapeutic coherence, and the lack of open-ended dialogue meant that extended use could lead to repetitive responses and diminished emotional resonance.

Previous SIHP trials that utilised a personalised VR avatar and a web-based chatbot yielded positive results \cite{RefWorks:edalat2025self-initiated}. However, switching between these platforms disrupted affective flow. Furthermore, the rule-based chatbot lacked open-ended dialogue, resulting in repetitive interactions and fading emotional resonance during extended use.

\subsection{Mobile Mental Health: From Rule-Based Bots to LLMs}
\label{sec:mHealth}

Mobile mental health interventions have often been built around structured psychotherapy frameworks delivered through text-based, guided workflows. Early applications such as Woebot \cite{woebot2026official} and Wysa~\cite{wysa2026official} demonstrated the feasibility of digital interventions, with randomised controlled trials showing significant reductions in depression symptoms \cite{fitzpatrick2017delivering}. However, evaluations consistently highlight the repetitive nature of scripted, decision-tree systems and their inability to handle nuanced inputs. With the emergence of large language models (LLMs) such as ChatGPT \cite{openai2023chatgpt} and Gemini~\cite{team2023gemini}, research has shifted toward generative dialogue for more flexible interaction.

Yet more human-like conversation does not automatically yield safe or professional therapeutic communication. Scholich et al. \cite{scholich2025comparison} compared general-purpose LLM chatbots with licensed therapists and found systematic differences: therapists support deeper engagement through evoking elaboration and sustained inquiry, whereas LLMs tend toward shallow interaction marked by excessive reassurance and premature advice-giving. Wang et al. \cite{wang2025evaluating} found that an LLM engineered to deliver cognitive restructuring (CR) could adhere to core CR protocols and provide empathic validation, but introduced subtle power imbalances. In high-risk contexts, Cui et al. \cite{cui2025development} developed a self-help suicide-intervention LLM chatbot reporting strong perceived usability, emotional support, efficacy, and safety. Such results underscore that higher-risk contexts demand explicit safety guardrails and auditable processes. These findings indicate that LLMs should be embedded within clearly specified therapeutic protocols rather than deployed as general-purpose agents.

\subsection{Empathy, Mirroring, and Therapeutic Support}
\label{sec:Empathy}

The limitations of LLMs in therapeutic depth often stem from a lack of embodiment. In traditional therapy, clinicians build a therapeutic alliance not through verbal content alone but through psychological inference and synchronised non-verbal behaviour. Empathy is the bedrock of this alliance, enabling understanding of another's experiences, and comprises person, cognitive, and affective empathy~\cite{bohart1997empathy,watson2016role}.

Mirroring, in which the therapist imitates the client's non-verbal cues such as facial expressions and actions \cite{knol2020reformulating}, has a bidirectional association with affective empathy \cite{HENDRIKSE2023138}: more empathetic people mimic others' facial expressions more consistently \cite{dimberg2011emotional}. Mirroring a client's movements helps them feel acknowledged and validated, encouraging them to share feelings and strengthening the working relationship. Hendrikse et al. \cite{HENDRIKSE2023138} provide a framework for incorporating mirroring in virtual environments through synchronised facial expressions and body posture, modelled with two virtual agents that decipher the client avatar's non-verbal cues and mirror them in the therapist avatar. However, their expressions ranged only from neutral to positive, and the realism of the actions and emotions was not verified by human participants.

\subsection{Expressing Emotion: From Basic Emotions to Embodied Cues}
\label{sec:emotion-categories}
Ekman's model of basic emotions is the most widely recognised framework for emotion research, defining emotions as innate, biologically based responses found across cultures and species, in contrast to culturally variable secondary emotions \cite{ekman1992there}. Ekman identifies six core emotions—anger, disgust, fear, happiness, sadness, and surprise—and notes that an observer's estimate of emotion is safest when read as a combination of facial, skeletal, vocal, autonomic, and coping responses. Secondary emotions incorporate situational context to display mixtures of basic emotions; Becker-Asano \cite{becker2008affect} terms these \enquote{adult} emotions (e.g. relief, hope) as they arise from higher cognitive processes.

Non-verbal tools such as facial expressions and body gestures are key to expressing emotion and more universal than speech or text \cite{leong2023facial}, with facial expressions in particular showing relative universality across geographic and cultural backgrounds. To model facial expressions, the Facial Action Coding System (FACS) \cite{ekman1978facial} is among the most popular systems \cite{tolba2018realistic}, coding individual facial muscles as action units (AUs). Body movements provide an additional source of emotional information and modulate emotion conveyed through the face and voice \cite{aviezer2008angry}. Their prototypical form varies by culture \cite{dael2012emotion}: collectivistic cultures value social harmony over self-expression, yielding more regulated, subtler body movements than individualistic cultures \cite{goudbeek2020cultural}. Hossain et al. \cite{hossain2025exploring} have explored modelling this in robots to evaluate a robot's capability for empathetic behaviour in human-robot collaboration. Unlike \cite{HENDRIKSE2023138} which looks at affective empathy, they model empathy through a combination of cognitive empathy and a corresponding affective response, coining it as empathic behaviour. For example, if the user looked unhappy, the robot would infer that they were sad and would respond by nodding while patting their back.

Speech synthesis and recognition technologies bridge human affect and machine processing \cite{sudhan2024texttospeech,dixit2026comparative}. Applications such as Woebot~\cite{woebot2026official}, Wysa \cite{wysa2026official}, Youper \cite{youper2026official} and Replika \cite{replika2026official} incorporate TTS to speak with human-like voices, fostering companionship and perceived empathy while improving accessibility for users who prefer auditory interaction \cite{sudhan2024texttospeech}. Beyond conversational agents, Calm \cite{calm2026} uses vocal guidance for relaxation, breathing, and bedtime stories, illustrating how voice supports scalable, low-friction emotional regulation.

\section{Design and Implementation}
We developed an iOS prototype that delivers SAT/SIHP through an \emph{embodied empathy} paradigm. The application features a multimodal architecture, integrating a prompt-engineered LLM with 3D avatars and AR to refine the core therapeutic experience within a standardised ecosystem.

% We developed a mobile application that delivers SAT/SIHP through an \emph{embodied empathy} paradigm. For this initial prototype, the application was built for iOS, allowing us to refine the core therapeutic experience within a standardised ecosystem before extending the application to other platforms. The application adopts a multimodal architecture that integrates a prompt-engineered LLM with the expressive visual presence of 3D avatars and AR. %We have created a digital ecosystem grounded in self-attachment theory, designed to transition psychotherapy from a text-based exercise into a continuous, embodied experience. 

%This section outlines the application's architecture, describing the design rationale and technical implementation of each stage.

\subsection{Application Overview}
\label{sec:user_flow}

\begin{figure}[H]
    \centering
    \includegraphics[width = 0.85\hsize]{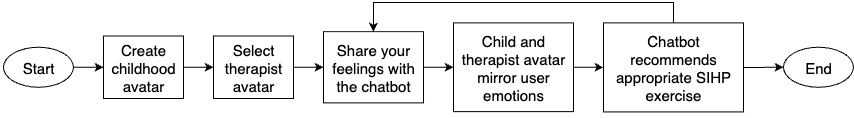}
    \caption{Flowchart of user interaction journey from onboarding and avatar creation to guided dialogue.}
    \label{fig:UserFlowDiagram}
\end{figure}

\subsection{The Multimodal Therapist Avatar System}

\noindent{\bf Avatar Creation and Display}
Users can create a personalised childhood avatar and select a therapist from a diverse gallery (Figure~\ref{fig:avatarcreate}). This design is grounded in the \enquote{self-compassion} concept of the SAT/SIHP protocol, where externalising emotions to a representation of one's childhood self facilitates emotional reconciliation. By providing customisation, we aim to increase the user's sense of ownership and agency, fostering a deeper emotional connection than using generic placeholders. The application integrates the MetaPerson Creator \cite{metaperson_loader_unity} via a UniWebView overlay, enabling users to generate high-fidelity 3D avatars from personal photos. These models are stored as GLB file links in a cloud database \cite{cloudflare2025d1} and retrieved using 8-digit unique share codes (Figure~\ref{fig:app_screenshots}). For therapist selection, a carousel-based UI allows users to browse pre-computed avatars, ensuring a diverse range of visual identities that meet different user preferences for representation.

% \begin{figure}[H]
%     \centering
    
%     \begin{subfigure}[b]{0.3\textwidth}
%         \centering
%         \includegraphics[width=0.69\textwidth]{figures/chatroom.PNG}
%         \caption{Session Initialisation}
%         \label{fig:chatroom}
%     \end{subfigure}
%     \begin{subfigure}[b]{0.3\textwidth}
%         \centering
%         \includegraphics[width=0.69\textwidth]{figures/chat_4.PNG}
%         \caption{Positive Affective Mirroring}
%         \label{fig:emotion_render_pos}
%     \end{subfigure}
%     \begin{subfigure}[b]{0.3\textwidth}
%         \centering
%         \includegraphics[width=0.69\textwidth]{figures/chat_6.PNG}
%         \caption{Negative Affective Mirroring}
%         \label{fig:emotion_render_neg}
%     \end{subfigure}

\begin{figure}[t]
    \centering
    \begin{subfigure}[b]{0.32\columnwidth}
        \centering
        \includegraphics[width=\linewidth]{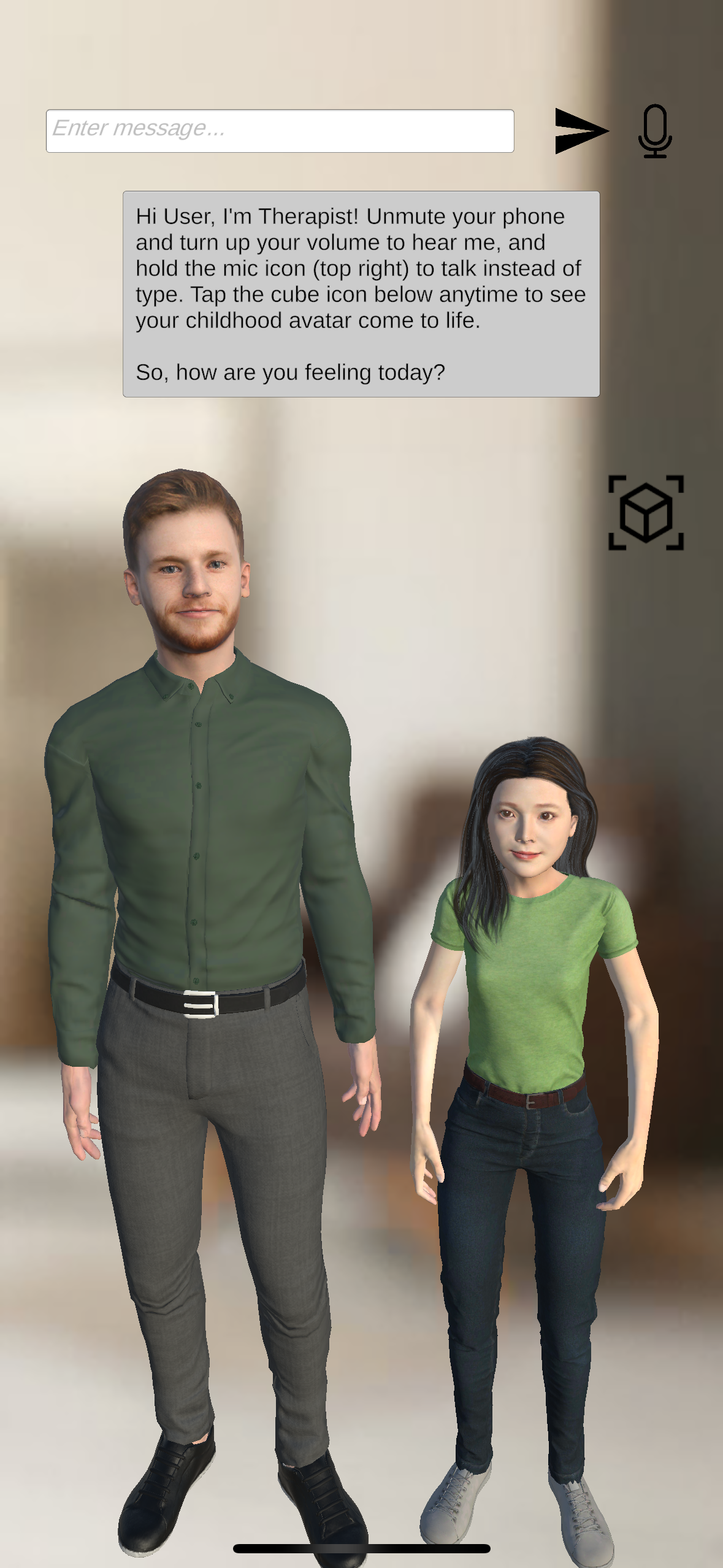}
        \caption{Session Initialisation}
        \label{fig:chatroom}
    \end{subfigure}
    \hfill
    \begin{subfigure}[b]{0.32\columnwidth}
        \centering
        \includegraphics[width=\linewidth]{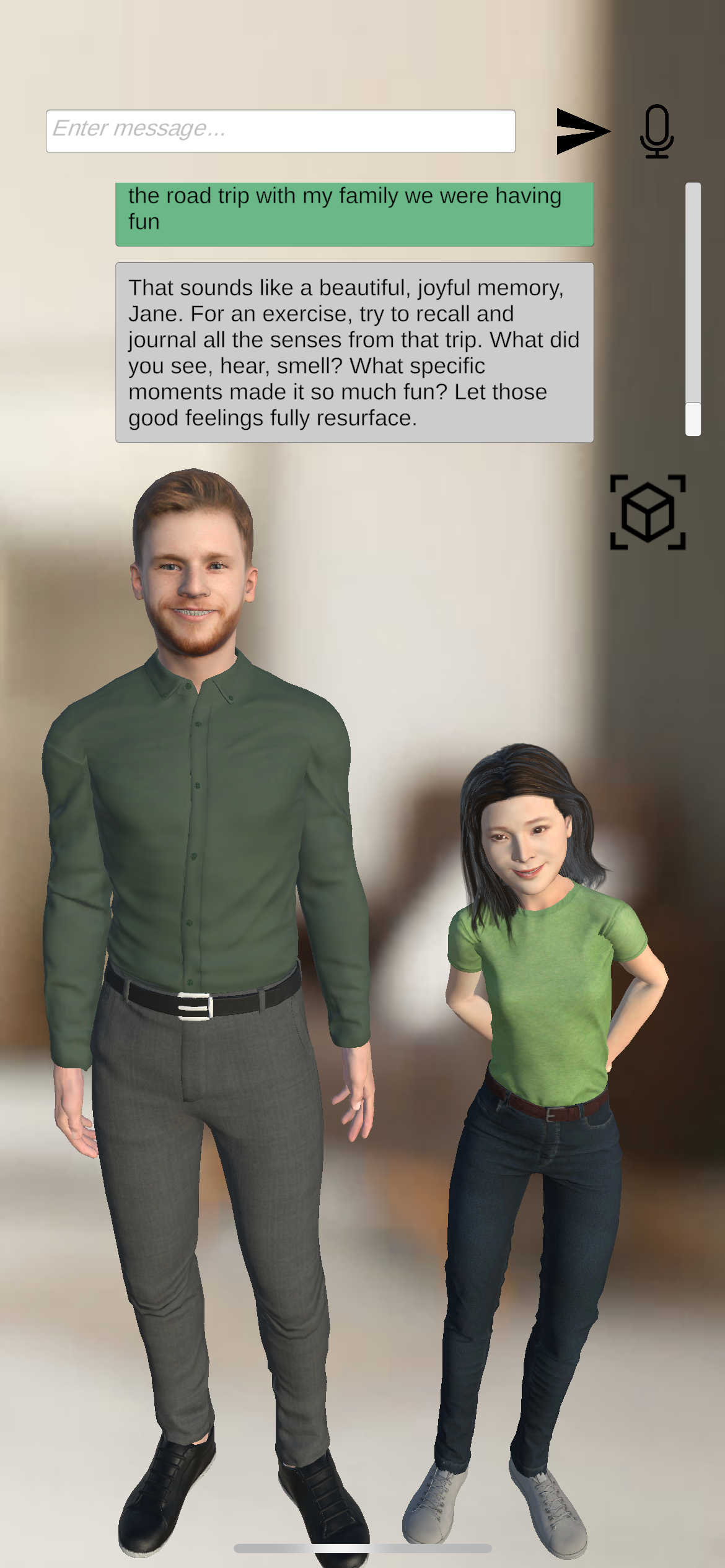}
        \caption{Positive Affective Mirroring}
        \label{fig:emotion_render_pos}
    \end{subfigure}
    \hfill
    \begin{subfigure}[b]{0.32\columnwidth}
        \centering
        \includegraphics[width=\linewidth]{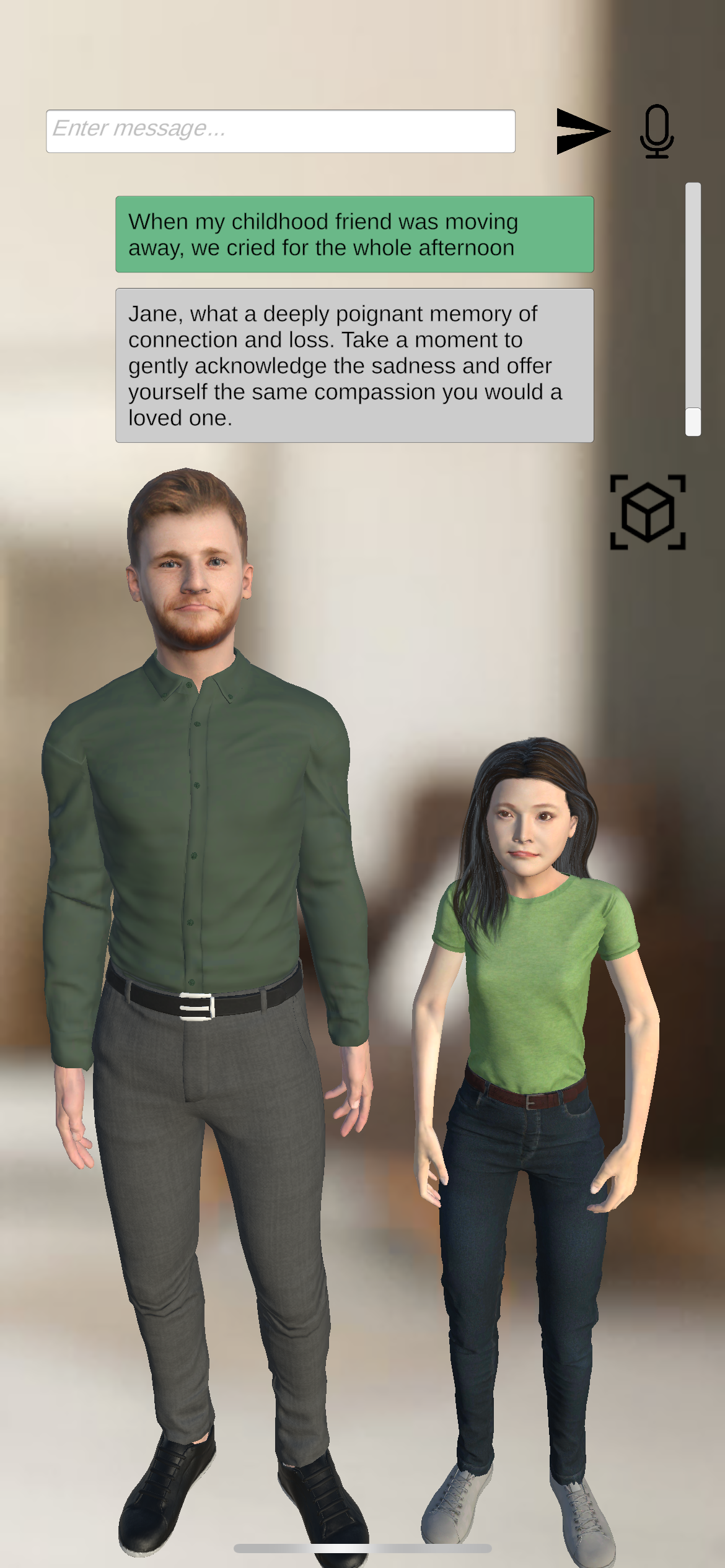}
        \caption{Negative Affective Mirroring}
        \label{fig:emotion_render_neg}
    \end{subfigure}

   \caption{Automated emotion mirroring within the Virtual Therapy Room. (a) Session Initialisation: the therapist chatbot opens with a welcoming greeting. (b) Positive and (c) Negative Affective Mirroring: both avatars transition to a ``Happy'' or ``Sad'' state in response to the emotion detected from user input.}

    \label{fig:app_screenshots2}
\end{figure}
%Xinyan: The above figure does not appear to be linked to the text. By linking it we can reduce the amount we put in the captions and reduce a lot of text - as below there is mention about things here
\noindent{\bf Chat Interface and Prompt-Engineered LLM}
These are designed to provide \enquote{mirroring and containment} — key psychotherapeutic techniques. As shown in Figure~\ref{fig:VTR}, the chat window is at the top of the screen, to avoid obstructing the therapeutic dialogue, thereby reducing cognitive load for users. The LLM's persona is prompt-engineered to be an empathetic listener rather than a rigid rule-based bot, facilitating a more natural and engaging therapeutic alliance. We utilised Gemini LLM \cite{team2023gemini}, optimised through an iterative \enquote{prompt science} framework \cite{shah2024prompt} to ensure the chatbot remains a supportive facilitator of the SAT/SIHP protocol. The final prompt instructs the model to validate user feelings (mirroring), reiterate and label the user's negative emotions, providing clarity to feelings that might feel vague or overwhelming. The workflow starts from the Chatbot module, which transmits user text to a cloud-hosted emotion recognition model. The resulting emotion classification is processed by the Emotion Rendering engine to drive the avatar display through animations.

% %Xinyan: This diagram doesn't seem to add much actually. You can just explain in the text as it is not complicated, so could go
 
% \begin{figure}[H]
% \centering
% \includegraphics[width = 0.55\hsize]{figures/OverallArchitectureDiagram.png}
% \caption{Integrated pipeline between the local Unity environment and Google Cloud services.}
% \label{fig:ArchitectureDiagram}
% \end{figure}

\noindent{\bf Emotion Recognition and Avatar Animation}
To investigate effective ways of \enquote{embodied empathy} in virtual therapy, we automated the transition from text input to visual emotional display. Instead of requiring manual selection of expressions, the system automatically mirrors the user's detected emotions in the child avatar, while the therapist avatar displays \enquote{affective empathy}—a regulated, calm response to negative emotions. This design choice aims to strengthen the user's emotional bond with their inner child through immediate, non-verbal feedback. The system leverages a cloud-hosted RoBERTa-based emotion recognition model \cite{liu2019roberta} to analyse user text in real-time (mean end-to-end latency $\approx$ 2s, encompassing network RTT and model inference). The model was double fine-tuned and achieved an accuracy of 94.96\% and a macro-F1 score of 95.10\% \cite{alazraki2021empathetic}. These classifications trigger specific FACS-based blendshape animations \cite{trioux2025exploring} (e.g., specific Action Units configured for Happy, Sad, Angry, Fearful, Disgusted, Surprised, Neutral) and Mixamo skeletal animations, as shown in Figure~\ref{fig:app_screenshots2}.
%While the child avatar mirrors the user's intensity through body movement, the therapist avatar is restricted to subtle facial expressions to maintain a sense of professional stability and psychological safety. 

\noindent{\bf Multimodal Interaction (STT/TTS)}
For the SAT/SIHP protocol, the introduction of voice interaction is driven by a critical observation: at moments of peak emotional arousal, the act of typing becomes a significant cognitive distractor. Traditional text entry demands complex fine motor control, syntactic organisation, and continuous visual-feedback monitoring. For a user in a state of distress, these requirements create interaction friction that can interrupt the spontaneous flow of thoughts and feelings. We hypothesised that voice would enable a more fluid and natural disclosure process when users are emotionally activated, reducing friction compared to text-only interaction. We also sought to minimise cognitive dissonance and increase therapist presence by matching the synthesised voice (e.g., perceived gender) to the selected therapist avatar, thereby making the interaction feel more like a human-to-human exchange. We integrated Google Cloud STT and TTS APIs \cite{googlecloudstt2026,googlecloudtts2026} within the Unity environment \cite{unitytechnologies-unity}. Users can toggle a \enquote{press-and-hold} microphone icon to record speech, which is then transcribed and sent directly to the LLM. All messages returned from Gemini are displayed on the screen and the TTS API is automatically called to read the text out loud. To ensure cross-modal consistency of the therapist, the TTS engine automatically selects a voice tone that strictly matches the gender and visual persona of the selected therapist avatar. This alignment of visual and auditory identities is critical for preserving immersion and conversational continuity.

%To investigate the effect of voice-enabled interaction on session engagement, we included speech-to-text (STT) and text-to-speech (TTS) features. 

\noindent{\bf AR Mode: Spatial Co-Presence}
To enhance spatial presence and co-presence, we implemented an AR modality that projects the childhood avatar directly into the user's physical environment. Our design hypothesis suggests that observing one's childhood self standing in a familiar, real-world context serves as a powerful grounding mechanism, thereby amplifying the emotional resonance of the SAT/SIHP exercises and strengthening user-avatar attachment. While our initial concept envisioned a continuous AR background during chatbot interactions, preliminary testing revealed critical ergonomic and experiential constraints: First, maintaining a tracking-permissive device posture while typing caused physical fatigue. Second, users reported symptoms of motion sickness during prolonged sessions, as the constant camera movement conflicted with the static UI elements. Consequently, we refined the feature into an optional, on-demand modality (Figure~\ref{fig:armode}). The feature was built using Unity's AR Foundation \cite{arfoundation2026} and Apple ARKit \cite{arkit2026}.

\section{Evaluation}

Our overarching research question is whether personalised avatars, AR interaction, and emotion mirroring can enhance engagement with SAT/SIHP. To assess the feasibility and user acceptance of the mobile application, we conducted an eight-day user study. %Ethical approval for the study was obtained from the institution.

\subsection{Study Design and Participants}
We recruited non-clinical, English-speaking adults to use their personal devices in naturalistic settings for 20–45 minutes daily over eight consecutive days. The protocol was structured into two phases to systematically probe the impact of different interaction modalities:

 %This phase focused on the fundamentals of SAT.
 %This phase focused on practicing and learning SIHP exercises daily.

\noindent{\bf Phase 1: Self-Attachment (Days 1–2)}:  Participants created a 3D child avatar from a personal childhood photo, representing their childhood self. They were then guided to interact with this child avatar in a virtual therapy room and in AR. The goal was to establish an initial emotional and empathic bond with their childhood self.

\noindent{\bf Phase 2: SIHP Exercises (Days 3–8)}:  Participants engaged with the LLM-driven therapist and their child avatar in the virtual therapy room or in AR to complete non-hostile humour tasks. These exercises were intended to help them apply SIHP principles to everyday emotional experiences.

\subsection{Data Collection and Measures}
We collected both quantitative and qualitative data through a post-study Qualtrics questionnaire comprising:
%The list of question items can be found in Appendix \ref{appendix:questions}. 

\begin{tight_itemize}
  \item 30 rating statements on a 5-point Likert scale (1 = Not at all to 5 = Extremely),
  \item 2 multiple choice questions assessing perceived mood impact,
  \item 4 open-ended questions exploring positive/negative aspects, missing features, and suggestions for improvement.
\end{tight_itemize}

To facilitate comparison across our multimodal application, we normalised the Likert responses to a 0--100 scale and aggregated them into three composite dimensions:

% Because our mobile application integrates multiple modalities within a single therapeutic journey, individual questionnaire items capture partially overlapping aspects of experience. To facilitate comparison across application components, we aggregated items into three composite dimensions. Specifically, we normalised all Likert responses to a 0--100 scale and grouped them as follows:

\begin{enumerate}
  \item \textbf{Avatar Interaction}: Perceived similarity, realism, and satisfaction with the child avatar; satisfaction with therapist avatar options, and the perceived usefulness of AR-based interaction.

  \item \textbf{Chatbot and Multimodal Efficacy}: Clarity, helpfulness, and empathy of Gemini-driven responses; the intuitiveness and helpfulness of STT and TTS features.

  \item \textbf{UX and Therapeutic Engagement}: General usability (ease of use, layout, efficiency), perceived glitches and recovery; the extent to which the app and its features supported emotional engagement and motivation to complete the exercises.

\end{enumerate}

%Xinyan: Maybe this paragraph can be removed as discussion covers all these elements
%Sixteen participants completed the full eight-day protocol and the post-study questionnaire. Upon collection, normalised scores across all three dimensions fell in the mid-50s to high-60s, consistently above the neutral midpoint. All participants reported that the exercises had a positive impact on their mood, fourteen out of sixteen indicated that they would recommend the app to someone who needed it. These findings suggest that, for a motivated non-clinical population, the mobile application is both feasible and acceptable as a tool for delivering SAT/SIHP.

\section{Discussion}
Our goal was to investigate how different interaction modalities in a mobile SAT/SIHP application shape users' therapeutic experience. We analyse and discuss the results in
the following sections.

\begin{figure}[t]
    \centering
    \includegraphics[width=0.8\columnwidth]{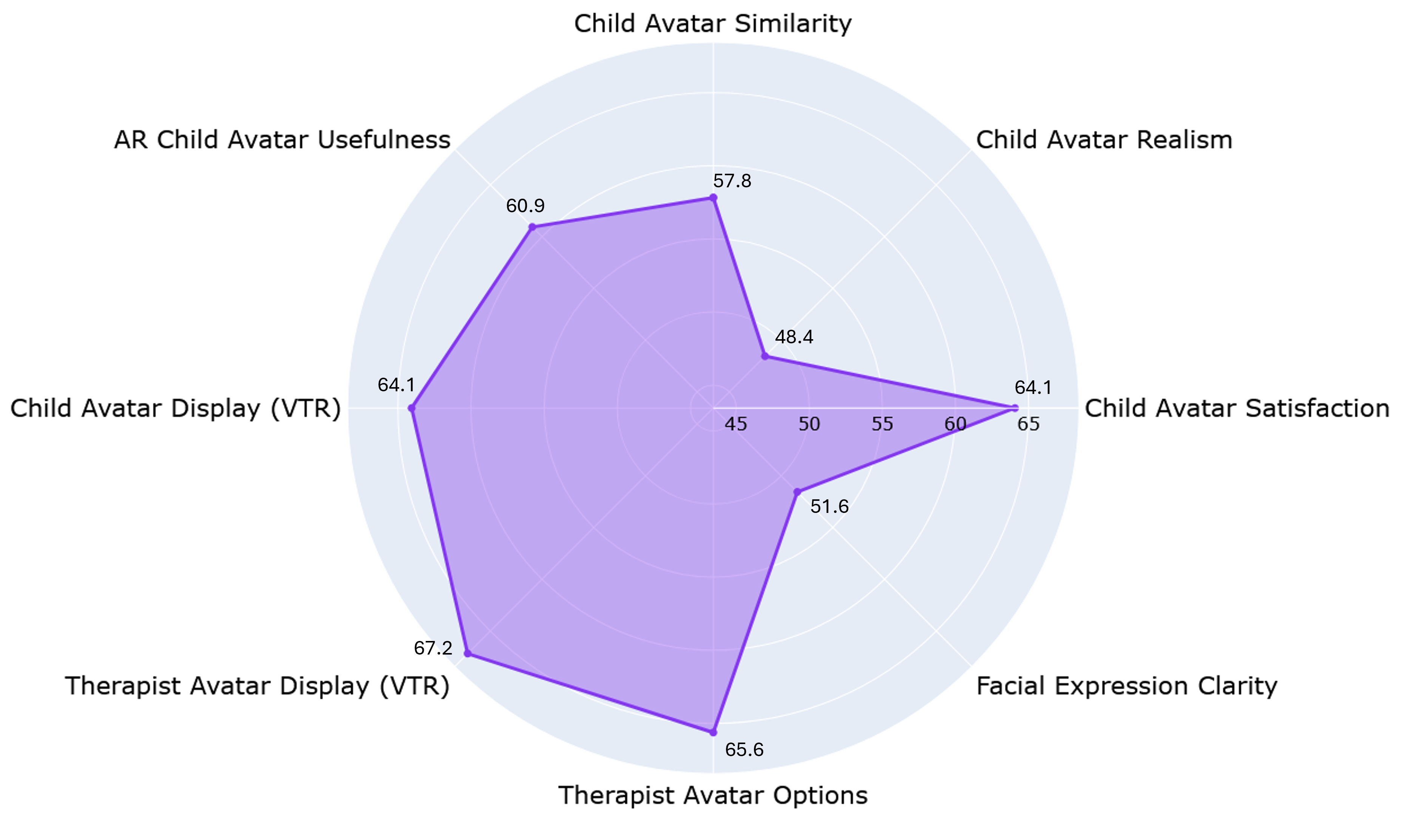}
    \caption{Avatar Mean Rating Radar Chart ($N=16$) across eight metrics in the Avatar Interaction dimension.}
    \label{fig:radar_avatar}
\end{figure}

\begin{table}[t]
    \centering
    \caption{Avatar cluster dimension scores ($N=16$): for visual and spatial components.}
    \label{tab:avatar_dims}
    \scriptsize
    \begin{tabular}{>{\raggedright\arraybackslash}p{3.2cm}cc}
        \toprule
        \textbf{Dimension} & \textbf{Mean} & \textbf{SD} \\
        \midrule
        Child Avatar Similarity         & 57.8 & 21.8 \\
        Child Avatar Realism            & 48.4 & 19.3 \\
        Child Avatar Satisfaction       & 64.1 & 22.3 \\
        Facial Expression Clarity       & 51.6 & 23.2 \\
        Therapist Avatar Options        & 65.6 & 22.1 \\
        Therapist Avatar Display (VTR)  & 67.2 & 23.7 \\
        Child Avatar Display (VTR)      & 64.1 & 27.3 \\
        AR Child Avatar Usefulness      & 60.9 & 30.2 \\
        \bottomrule
    \end{tabular}
\end{table}

\subsection{Custom Child Avatars and AR for Emotional Bonding}
The findings suggest that personalised avatars meaningfully support engagement and bonding. Within the Avatar Interaction dimension in Figure~\ref{fig:radar_avatar}, ratings for satisfaction with the child avatar's presence, therapist avatar options, and having both avatars visible alongside the chat all clustered in the upper range (50–70/100). Participants described the avatar experience as \enquote{fascinating}, \enquote{very immersive}, and \enquote{much more engaging compared to the chatbot-only} interface. Several participants emphasised that seeing their child avatar's mood change in response to the conversation made them want to keep talking to the chatbot, aligning with SAT's goal of strengthening attachment to one's childhood self. However, there was a notable drop in scores for avatar realism and facial expression clarity, reflecting two primary technical constraints. First, the generation API is heavily dependent on input photo quality: restricted by the older, black-and-white, and low-resolution childhood photos uploaded by participants, some generated avatars struggled to achieve a realistic likeness. Second, the avatars' facial animations relied on predefined, static emotion configurations rather than fluid movements. This lack of dynamic expressivity made it difficult for the avatars to capture and convey the nuanced, real-time affective shifts of natural human conversation.

The AR mode also received generally positive evaluations. Ratings for the helpfulness of seeing the child avatar in an AR environment were above the midpoint, and participants reported that it was \enquote{interesting and helpful} to place the avatar in their physical surroundings. However, qualitative feedback highlighted cognitive and ergonomic overhead. Participants reported confusion about scale (\enquote{confused at first about how big the AR model was}) and uncertainty about \enquote{how to interact meaningfully with the avatar}. Others expressed that they expected more in-app guidance and requested richer AR interaction features. 

Taken together, these results indicate that custom child avatars are central to fostering an emotional connection with one's childhood self, while AR can amplify this connection in short, intentional episodes. To fully realise the potential of AR in the therapeutic workflow, it must be carefully designed and accompanied by supportive guidance that helps users understand when and how to use AR in ways that meaningfully enrich SAT/SIHP practice.

\subsection{Emotion Mirroring and Embodied Empathy}

We hypothesised that emotion mirroring in the child and therapist avatars would enhance emotional connection and empathy. The results suggest that mirroring is powerful but fragile: when well-attuned, it significantly deepens engagement, but when misaligned, it risks breaking the therapeutic illusion. 

Items targeting emotion recognition and empathy (e.g., consistency between the child avatar's expression and the user's self-reported state, perceived empathy of the therapist avatar) achieved moderate-to-high scores. Participants noted that \enquote{the avatars' emotional responses felt appropriate and aligned well with how I was feeling} and that \enquote{after telling something to the chatbot about my mood, it correctly changed my avatar's mood, and it was really engaging to continue talking}. 
At the same time, a subset of users reported misattuned or exaggerated responses. One participant noted that the child avatar sometimes shifted into an intense fear/anxiety pose that felt \enquote{weird and unexpected}. Another observed that \enquote{avatars sometimes reacted when my input was neutral / had no emotional change}. While some found these moments amusing, they also recognised that for someone feeling low, seeing their childhood self panic or collapse when they are only mildly worried could be jarring and undermine the sense of safety that SAT/SIHP relies on. Perceptions of visual realism showed similar ambivalence. Some participants wished for more photorealistic and kind-looking therapist avatars, whereas others remarked on uncanny valley effects and cautioned that faces that are too realistic might be uncomfortable in a therapeutic setting. These tensions highlight a key design implication: in self-attachment digital therapy, both visual realism and emotional intensity must be carefully moderated. Mechanisms to soften expressions, reduce intensity, or quickly correct misclassified emotions (e.g., a simple \enquote{This doesn't match how I feel} control) are likely to be critical for sustaining embodied empathy over time.

\begin{figure}[t]
    \centering
    \includegraphics[width=0.75\columnwidth]{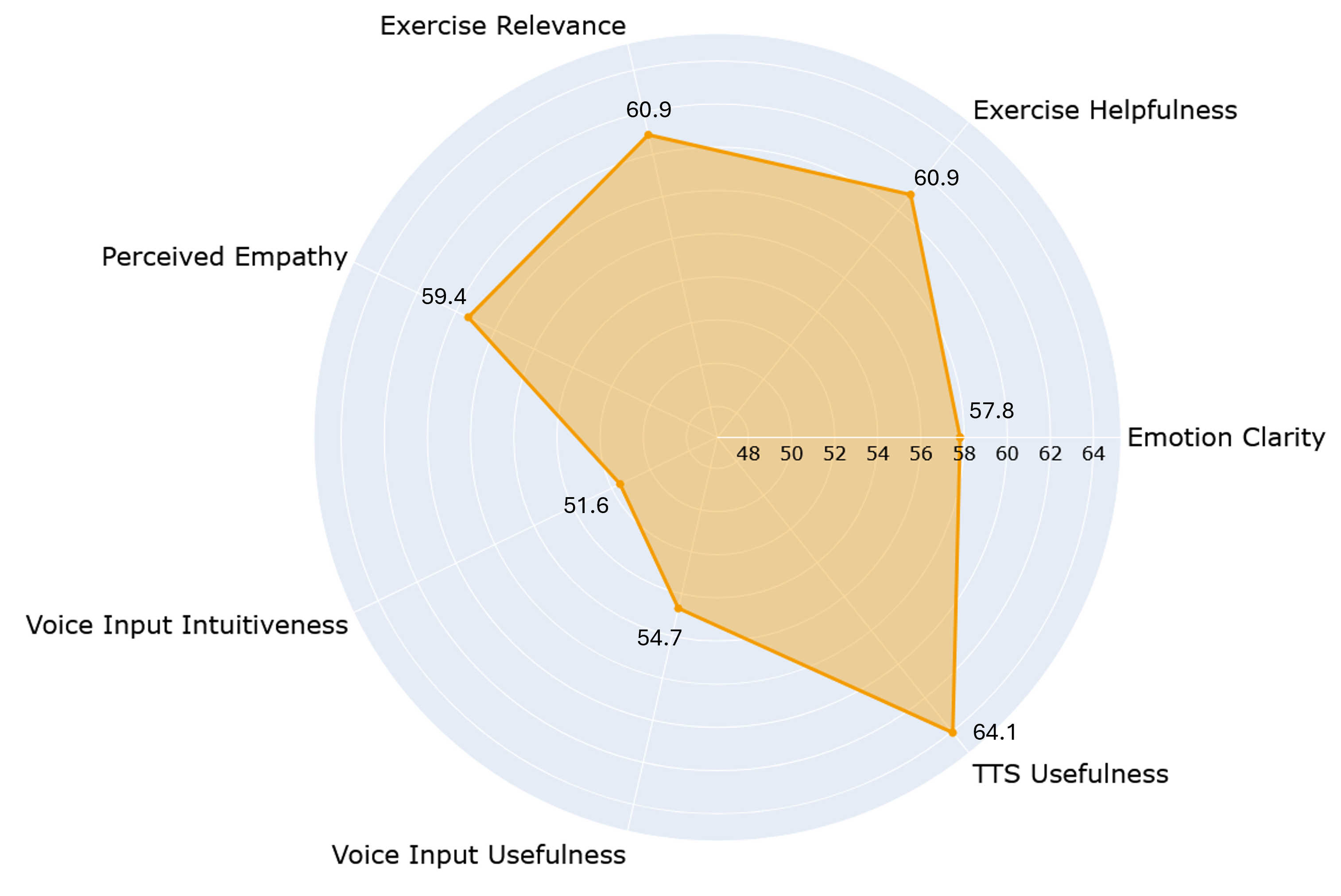}
    \caption{Chatbot \& Multimodal Radar Chart ($N=16$) across five metrics in the Multimodal Chatbot dimension.}
    \label{fig:radar_chatbot}
\end{figure}

\begin{table}[t]
    \centering
    \caption{Multi-Modal Chatbot scores ($N=16$) for LLM-driven responses and speech-based interaction features.}
    \label{tab:chatbot_dims}
    \scriptsize
    \begin{tabular}{>{\raggedright\arraybackslash}p{3.4cm}cc}
        \toprule
        \textbf{Dimension} & \textbf{Mean} & \textbf{SD} \\
        \midrule
        Emotion Clarity             & 57.8 & 25.4 \\
        Perceived Empathy           & 59.4 & 22.1 \\
        Voice Input Intuitiveness   & 51.6 & 29.5 \\
        Voice Input Usefulness      & 54.7 & 37.9 \\
        TTS Usefulness              & 64.1 & 27.3 \\
        \bottomrule
    \end{tabular}
\end{table}
 
\begin{figure}[t]
    \centering
    \includegraphics[width=\columnwidth]{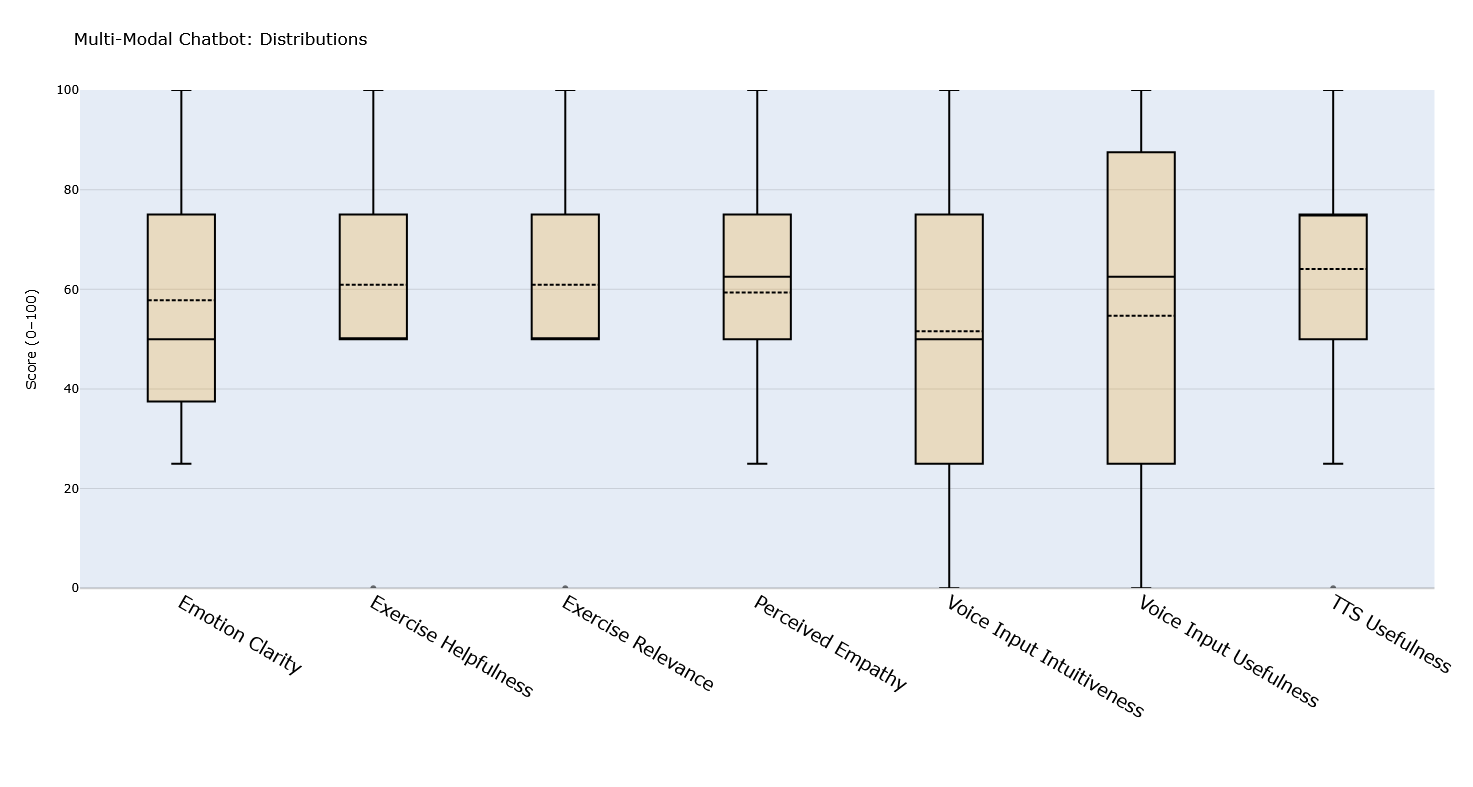}
    \caption{Distribution of participant ratings for the Multi-Modal Chatbot dimension.}
    \label{fig:box_chatbot}
\end{figure}

% High IQR in Voice Input Usefulness and Intuitiveness reflects diverse experiences with STT, whereas TTS Usefulness shows a more concentrated, higher-scoring distribution.

\subsection{Voice-Enabled Therapist and Expressive Practice}
Our third hypothesis concerned voice: we expected that a voice-enabled therapist would support more expressive practice during SAT/SIHP. The data show that voice output and input play different roles. As shown in Figure~\ref{fig:radar_chatbot}, the TTS feature was consistently well-received. Helpfulness of having chatbot responses \enquote{read} aloud was among the highest-scoring items in the Chatbot \& Multimodal dimension. Participants described the audio feedback as making the therapist feel \enquote{more immersive and empathetic} and the overall experience \enquote{refreshing}. TTS appears to support expression indirectly: by allowing users to listen rather than read, it reduces cognitive load and reinforces the feeling of being \enquote{spoken to} by a caring other, which aligns with containment goals in psychotherapy.

By contrast, STT produced a mixed picture. Figure~\ref{fig:box_chatbot} highlights this variance through high interquartile ranges (IQR) in the scores for Voice Input Intuitiveness and Usefulness. Some participants could not get the microphone working on their devices; others chose to type, stating that \enquote{it was fine to just type} or that they were \enquote{not used to it}. Even among those attracted to the idea of voice input, ratings for intuitiveness and helpfulness of STT were lower than those for TTS and text. Voice input also raises contextual concerns when participants use the app in shared or public spaces, where speaking aloud about sensitive issues may increase social exposure and perceived risk.\\

\begin{figure}[t]
    \centering
    \includegraphics[width=0.75\columnwidth]{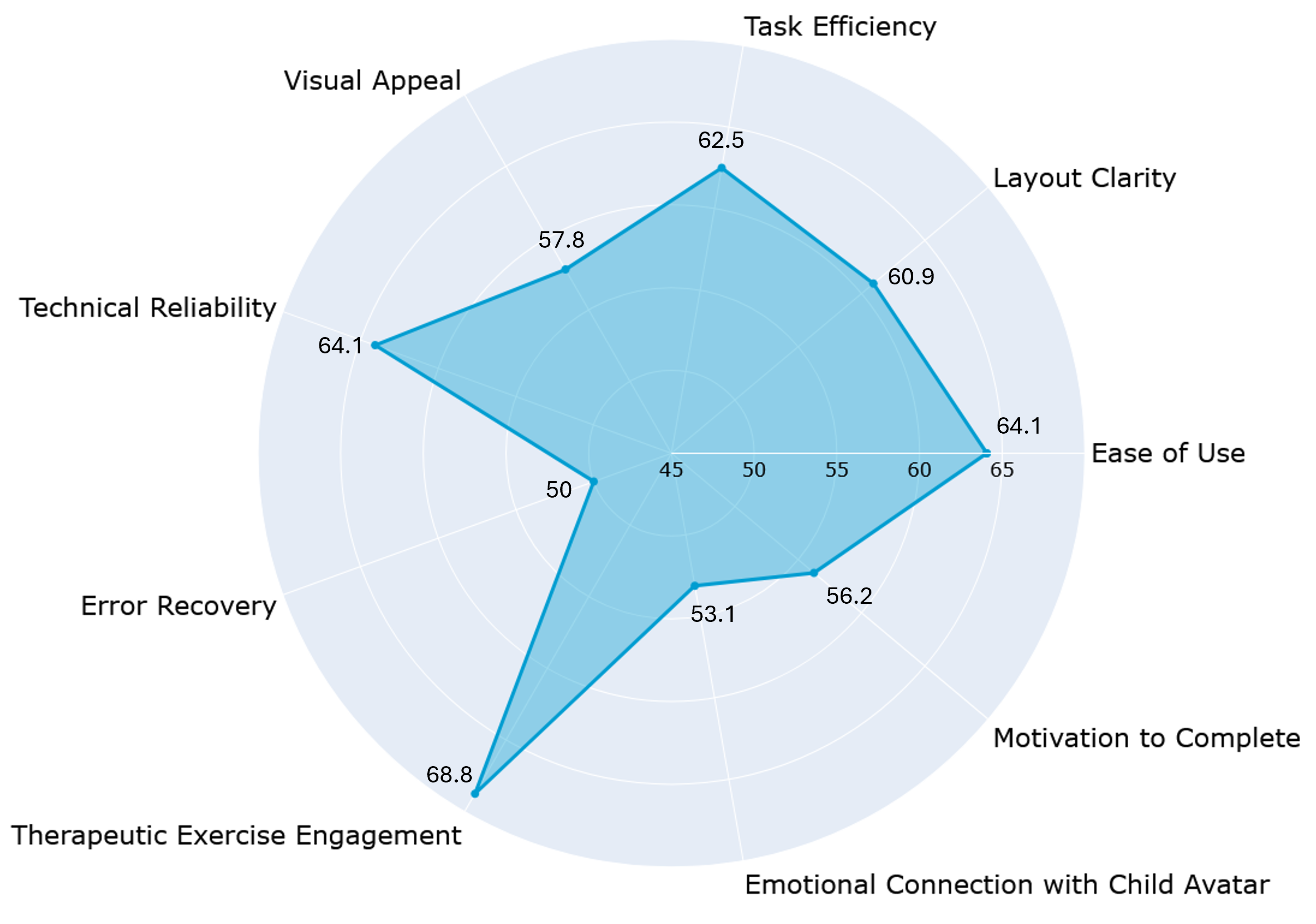}
    \caption{Therapeutic UX Radar Chart across nine metrics in the Therapeutic UX dimension.}
    \label{fig:radar_ux}
\end{figure}

\begin{table}[t]
    \centering
    \caption{Therapeutic UX cluster scores ($N=16$) for therapeutic experiences.}
    \label{tab:ux_dims}
    \scriptsize
    \begin{tabular}{>{\raggedright\arraybackslash}p{3.9cm}cc}
        \toprule
        \textbf{Dimension} & \textbf{Mean} & \textbf{SD} \\
        \midrule
        Ease of Use                             & 64.1 & 18.2 \\
        Layout Clarity                          & 60.9 & 18.2 \\
        Task Efficiency                         & 62.5 & 18.3 \\
        Visual Appeal                           & 57.8 & 23.7 \\
        Technical Reliability                   & 64.1 & 32.9 \\
        Error Recovery                          & 50.0 & 20.4 \\
        Therapeutic Exercise Engagement         & 68.8 & 23.3 \\
        Emotional Connection with Child Avatar  & 53.1 & 25.6 \\
        Motivation to Complete                  & 56.3 & 28.1 \\
        \bottomrule
    \end{tabular}
\end{table}

% \begin{figure}[H]
% \centering
% \includegraphics[width = 0.9\hsize]{figures/Box_UX.png}
% \caption{Distribution of participant ratings for the Therapeutic UX dimension. The distribution reveals robust systemic feasibility, with Therapeutic Exercise Engagement showing the highest overall concentration of positive scores. }
% \label{fig:box_ux}
% \end{figure}

\subsection{Beyond Modalities: Conversational Flow and Continuity}
Beyond the three hypotheses, the study revealed broader challenges that are critical for deploying SAT/SIHP in everyday life. First, participants wanted the therapist to carry more of the conversational burden. While many described individual responses as \enquote{thoughtful}, they also reported that the chatbot rarely asked follow-up questions or proactively suggested next steps, making the dialogue feel \enquote{a bit too robotic}. To realise the potential, future versions should integrate stronger state tracking and exercise recommendation.

The overall therapeutic user experience is illustrated in Figure~\ref{fig:radar_ux} and Table~\ref{tab:ux_dims}. We found that seemingly mundane UX decisions had a strong impact on the therapeutic frame. Participants expressed frustration at needing to remember share codes to reload avatars and at having to recreate avatars multiple times. In a protocol based on self-attachment, such friction is not merely inconvenient: repeatedly losing access to one's child avatar or therapist avatar disrupts the formation and maintenance of the intended bond. Future iterations should therefore prioritise robust avatar persistence and loading stability (e.g.,  secure on-device caching, biometric unlocking), with appropriate performance and privacy safeguards. Closely related to this need for system stability is the critical importance of graceful error recovery. Technical glitches like latency or API timeouts during vulnerable moments severely damage the therapeutic alliance. Resilient error handling, such as preserving user inputs during drops and providing empathetic fallback messages, is critical to maintaining continuity of care.

In summary, the study suggests that carefully orchestrated combinations of avatars, emotion mirroring, and vocalisation can make SAT/SIHP feel more engaging, expressive, and emotionally containing. At the same time, the realism of avatars and the intensity of emotional expression must be regulated, and the chatbot should take on more responsibility for guiding users through the therapeutic journey.

\section{Limitations and Future Work}
Several limitations constrain our findings. The sample was small (N=16) and comprised non-clinical, technologically literate adults who owned an iOS device. A cross-platform tool could broaden recruitment. The original SAT/SIHP course spans eight weeks to support gradual habit formation, whereas our condensed eight-day protocol compressed each week's content into a single day, leaving us without data on long-term adherence or engagement trajectories. Our measures captured subjective experience and self-reported mood rather than standardised clinical endpoints. Also, our data collection was limited by a lack of fine-grained interaction data, specifically, we did not log individual session durations or track users' real-time emotional states. Furthermore, confining the avatars' expressions to standard basic emotions restricted their range of expressivity, which may have fallen short of fully mirroring users' nuanced affective states. Finally, the underlying general-purpose LLM has not undergone clinical validation for hallucination and toxicity, which pose potential mental health risks. Our contribution should therefore be understood as mapping a design space for embodied, multimodal SAT/SIHP delivery and characterising early user experience, rather than demonstrating clinical efficacy.

Future work should conduct longer-term, more diverse trials evaluating how sustained interaction with child and therapist avatars affects wellbeing, attachment, and humour-related coping, combining system logs (e.g. interaction frequency, modality usage) with validated psychological measures. From a system perspective, future iterations should: (1) strengthen conversational statefulness and protocol-aware exercise recommendation (e.g. via long-term memory with Retrieval-Augmented Generation); (2) refine emotion mirroring with controllable intensity, mixed-emotion support, and user override; (3) improve persistent avatar access and loading via secure on-device storage; (4) adopt safety-tested, domain-specific LLMs with consent-based boundary-setting and harm-mitigation strategies \cite{b2024enabling}; and (5) develop an Android version. In the longer term, locally hosted open-source LLMs could enhance privacy and give designers finer control over safety and behaviour, provided they are carefully aligned with SAT/SIHP and evaluated under clinical supervision.

\section{Ethical Considerations}

%Xinyan: Here maybe you only need to talk about user study and the safety of the users, since the technical stuff is not relevant to the paper. I have commented the rest out. See what you think about it.
%Our mobile application prioritised generation speed, stability, and minimal data storage. P

Participants were fully informed about the use of LLMs, external APIs, and data handling before consenting to participate. The necessary ethical approval was obtained from the host institution. 

%For conversational functionality and TTS we used the Gemini API under a paid subscription. Dialogue histories are processed for the duration of a session and are not retained beyond that session. Data handling at the model endpoint is governed by Google's terms of service for API users \cite{googlecloudstt2026,googlecloudtts2026,team2023gemini}. The emotion classifier is deployed as a container on Google Cloud Platform and processes the current session's text to only infer an emotion label used for real-time avatar rendering. These labels are ephemeral and are not stored or shared outside the ongoing session. 

%For avatar creation, we store only the generated avatar link and an associated username in a Cloudflare D1 database \cite{cloudflare2025d1}, solely for saving and retrieving avatars within the app. Use of the original photo and the derived avatar is governed by the Itseez3D's terms of service \cite{avatarsdk-metaperson-api}, which state that the provider does not claim intellectual-property rights over user data. The avatars are designed for exclusive use within the application’s ecosystem. Their utility is restricted to the virtual therapy room and AR mode to reduce the risk of misuse. 

We recognise that the same technologies that enable supportive, empathic interactions could also be misused. For example, to create self-harming or harassing experiences \cite{chu2025illusionsintimacyemotionalattachment,b2024enabling}, or to foster unhealthy attachment to an AI agent \cite{moylan2025expert}. To mitigate these risks, our current design constrains the chatbot's role to SAT/SIHP-related coordination, avoids unsupported topics, and limits avatar functionality to therapeutic contexts. Future iterations will need stronger guardrails, including explicit boundaries in the dialogue, reminders of the system's artificial nature, and real-time moderation for unsafe or abusive content.

 \section{Conclusion}

In this work, we designed, implemented, and evaluated a novel multimodal mobile platform for delivering the Self‑Attachment Technique (SAT) enhanced with the Self‑Initiated Humour Protocol (SIHP). By unifying an LLM‑driven virtual therapist, customisable 3D avatars, and augmented reality within a single application, we transitioned SAT/SIHP from fragmented, rule‑based tools into an integrated, embodied therapeutic environment. Our eight‑day user study provides empirical evidence that a multimodal approach -- particularly the combination of personalised visual avatars and empathetic voice output—can effectively foster an affectional bond between users and their digital childhood selves. The study also revealed critical design trade‑offs: although automated emotion mirroring enhances perceived compassion, its success depends on high classification accuracy and carefully modulated intensity to avoid uncanny or disruptive experiences. Furthermore, participants expressed clear preference for proactive, context‑aware facilitation over reactive chatbot interaction, signalling that the next generation of AI‑supported therapy should guide rather than merely respond. 

%Future iterations will therefore benefit from integrating long‑term memory and Retrieval‑Augmented Generation (RAG) to enable more fluid, personalized, and coherent therapeutic dialogue.

As digital mental health increasingly adopts generative AI and immersive media, this research contributes a practical framework for designing systems that are not only accessible but also emotionally containing. By grounding technological innovation in the principles of self-attachment theory, we advance toward a new class of scalable digital companions—ones capable of delivering empathy, resonance, and therapeutic support on a global scale.

%%%%%%%%%%%%%%%%%%%%%%%%%%%%%%%%%%%%
\begin{acks}

We sincerely thank AvatarSDK for providing the API credentials for the MetaPerson Creator, Empowered Human Foundation for funding the user study deployment, and all participants for their time.

\end{acks}

%%%%%%%%%%%%%%%%%%%%%%%%%%%%%%%%%%%%
% REFERENCES
% ACM uses a specific bibliographic style. 
% You need a .bib file (e.g., references.bib)  
\bibliographystyle{ACM-Reference-Format}
\bibliography{references.bib} 

%%%%%%%%%%%%%%%%%%%%%%%%%%%%%%%%%%%%
\appendix
\section{Avatar Creation UI}
\label{appendix:avatar_UI}

\begin{figure}[H]
    \centering
    
    \begin{subfigure}[b]{0.2\textwidth}
        \centering
        \includegraphics[width=\textwidth]{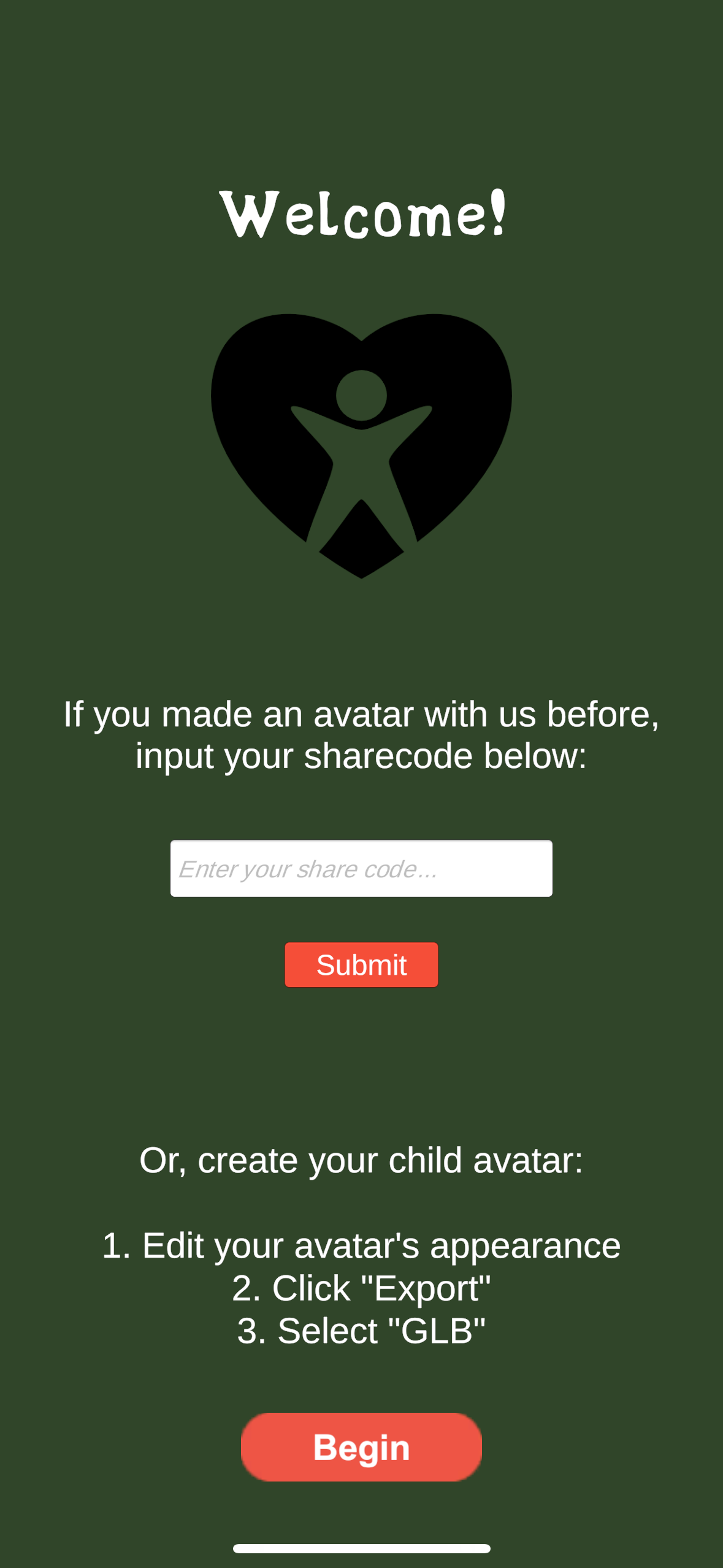}
        \caption{Homepage}
        \label{fig:homepage}
    \end{subfigure}
    \begin{subfigure}[b]{0.2\textwidth}
        \centering
        \includegraphics[width=\textwidth]{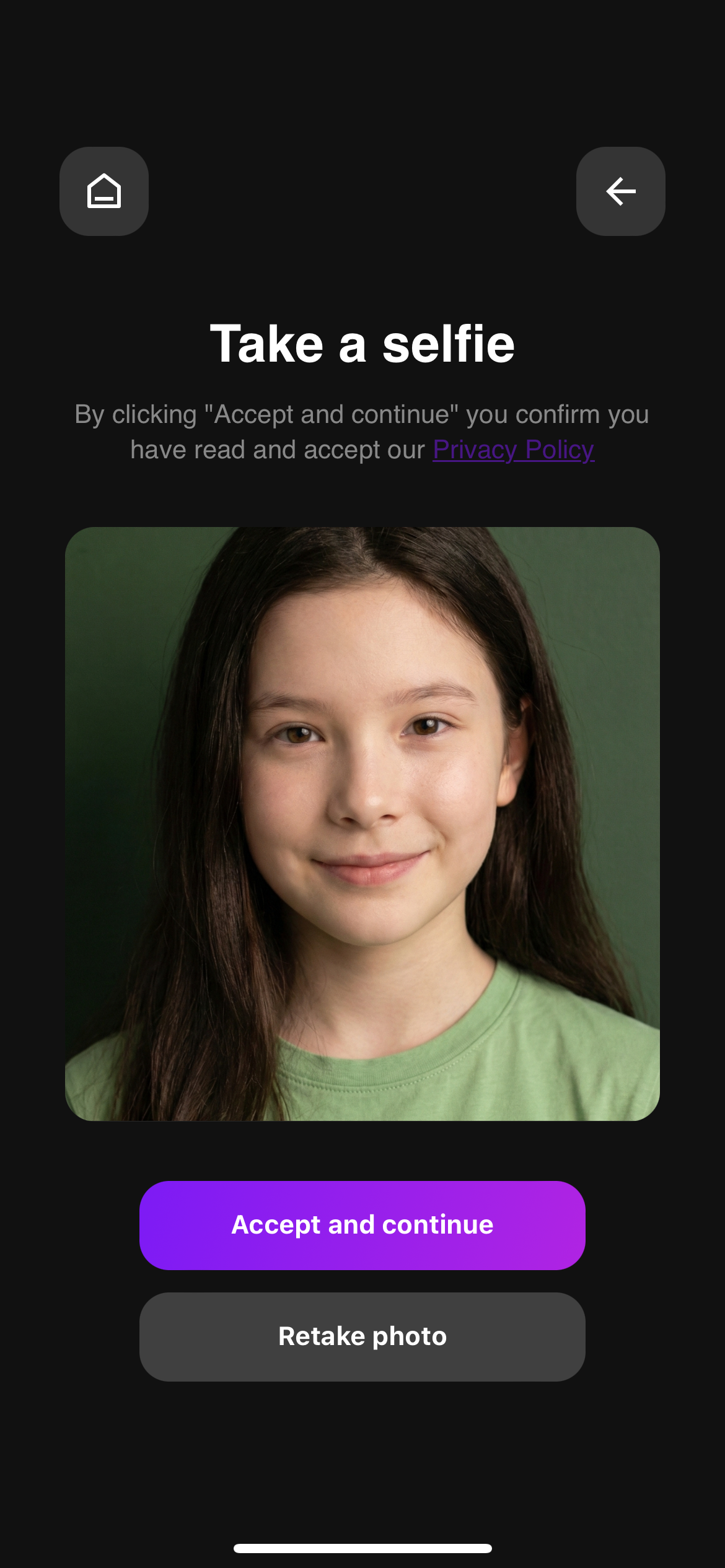}
        \caption{Create Avatar}
        \label{fig:avatarupload}
    \end{subfigure}
    \begin{subfigure}[b]{0.2\textwidth}
        \centering
        \includegraphics[width=\textwidth]{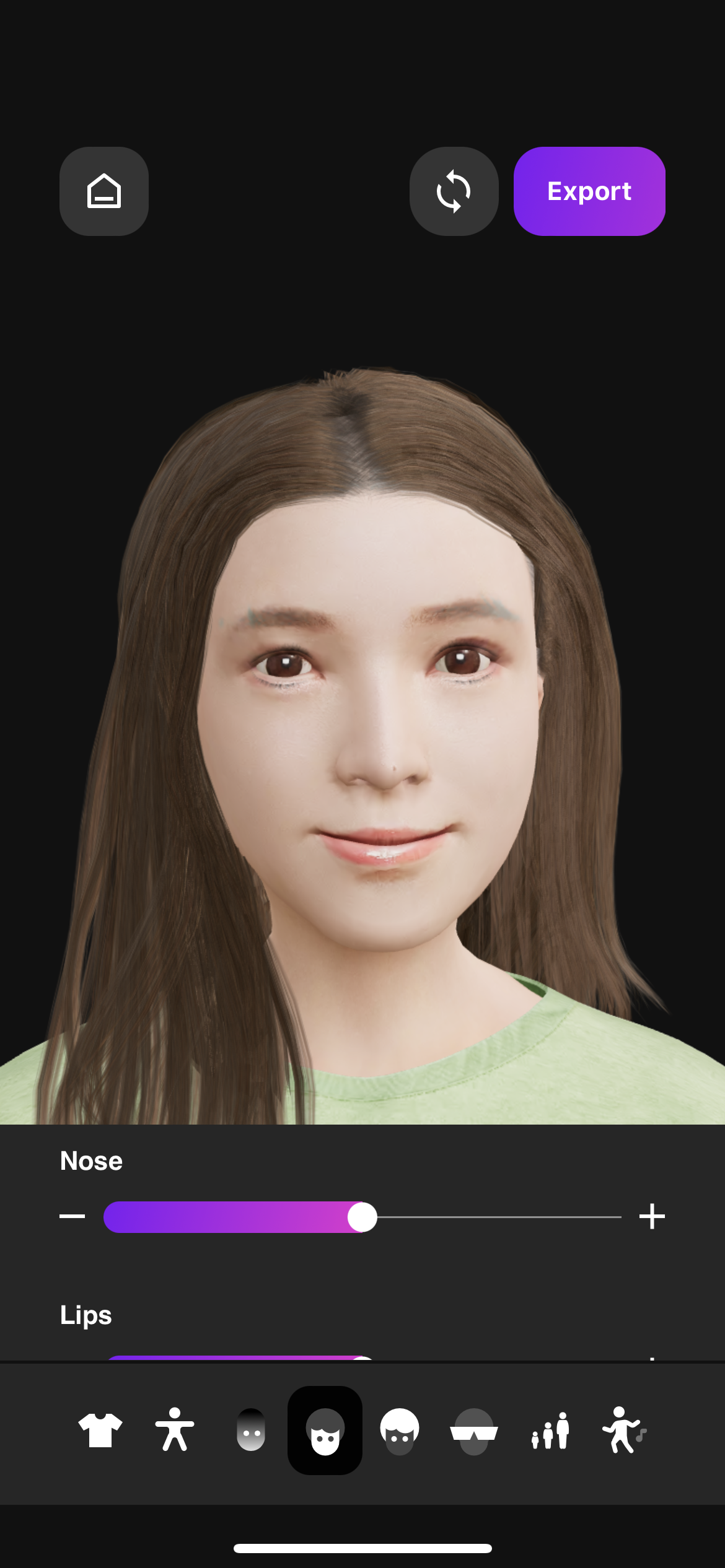}
        \caption{Customise Avatar}
        \label{fig:avatarcustom}
    \end{subfigure}
    \begin{subfigure}[b]{0.2\textwidth}
        \centering
        \includegraphics[width=\textwidth]{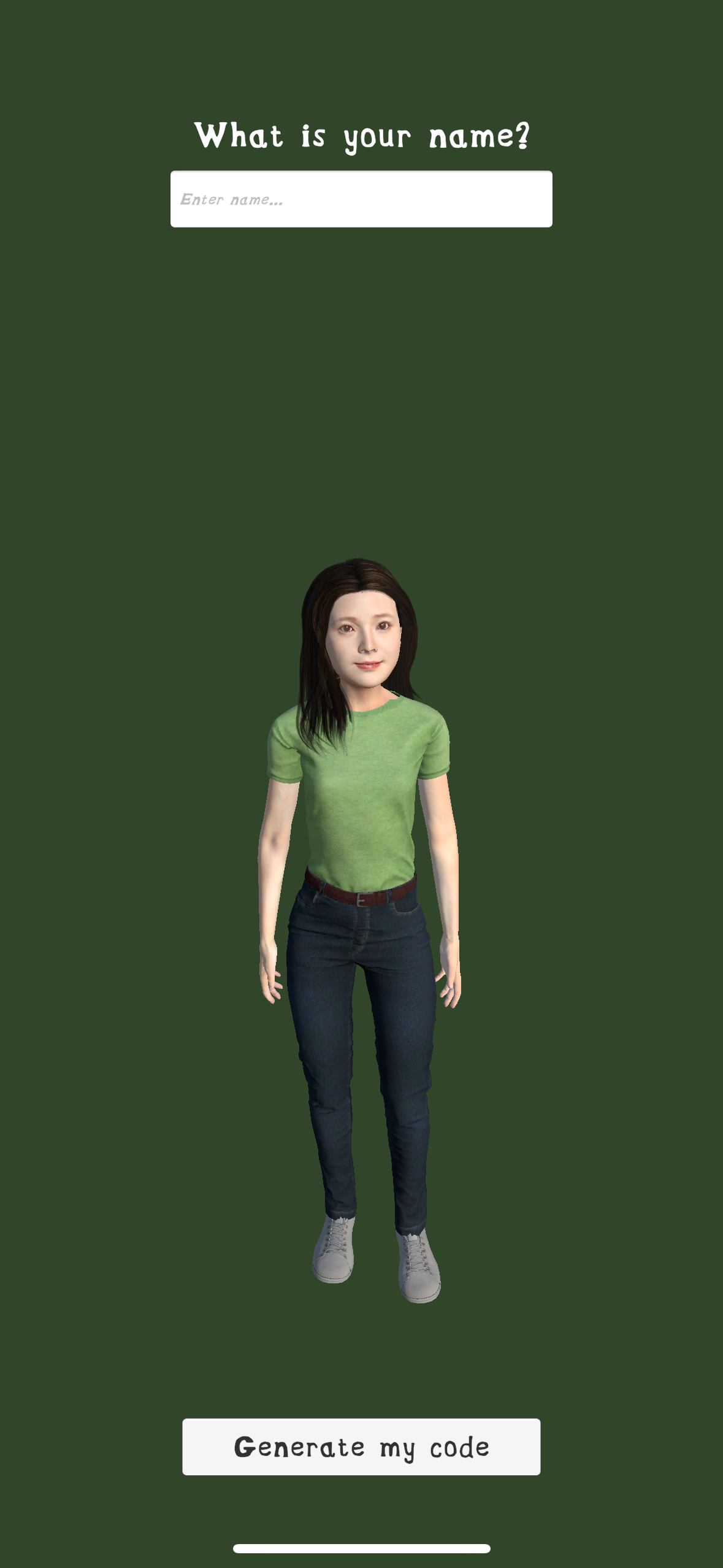}
        \caption{Save Avatar}
        \label{fig:avatarsave}
    \end{subfigure}

    \caption{Screenshots of the avatar creation phase of the application, where users create and customise their child avatar}
    \label{fig:app_screenshots}

\end{figure}

\section{User Study Questionnaire}
\label{appendix:questions}
\begin{table}[H] %  
\small % 
\centering
\caption{Mapping of Survey Items to Analytical Labels}
\label{tab:mapping}
\begin{tabularx}{\textwidth}{l p{3.5cm} X} % 
\toprule
\textbf{Category} & \textbf{Analytical Label} & \textbf{Survey Question / Statement} \\ 
\midrule

\textbf{Therapeutic UX} 
& Ease of Use & How easy did you find this application to use? \\
& Layout Clarity & How clear and well-organised do you think the app layout is? \\
& Task Efficiency & How quickly and efficiently were you able to complete the tasks? \\
& Visual Appeal & How visually appealing do you think the app is? \\
& Technical Reliability$^*$ & To what extent did you encounter any glitches or errors? \\
& Error Recovery & How easy was it to recover from a mistake made using the app? \\ 
\cmidrule{2-3}
& Exercise Engagement & Overall, how engaging was the experience of interacting with the avatars and chatbot? \\
& Emotional Connection & How emotionally connected did you feel to your avatars? \\
& Motivation to Complete & Did the app features help you stay motivated to complete the exercises? \\ 
\midrule

\textbf{Avatar} 
& Child Similarity & How similar does your child avatar appear compared to your childhood self? \\
& Child Realism & How realistic does the child avatar's appearance seem to you? \\
& Child Satisfaction & Overall, how satisfied are you with the final appearance of your child avatar? \\
& Facial Expression Clarity & How accurately do the child's facial expressions reflect your emotions? \\
& Therapist Options & How satisfied are you with the therapist avatar model options? \\ 
\cmidrule{2-3}
& Therapist Display (VTR) & How helpful was it to see the therapist avatar displayed alongside the interface? \\
& Child Display (VTR) & How helpful was it to see the child avatar displayed on the chatbot page? \\
& AR Child Usefulness & How helpful was it to see your child avatar in an AR environment? \\ 
\midrule

\textbf{Multi-Modal} 
& Emotion Clarity & How much clarity did you gain about your own emotions? \\
\textbf{Chatbot} 
& Exercise Helpfulness & How helpful were the exercises recommended by the chatbot? \\
& Exercise Relevance & How relevant were the exercises recommended by the chatbot? \\
& Perceived Empathy & How empathetic did the chatbot's responses feel to you? \\ 
\cmidrule{2-3}
& Voice Intuitiveness & How intuitive was the voice input (speech-to-text) feature? \\
& Voice Usefulness & How helpful was the voice input feature for expressing yourself? \\
& TTS Usefulness & How helpful was it to have the chatbot responses read aloud? \\ 
\midrule

\textbf{Qualitative /} 
& Feedback: Emotions & Were there any instances where the avatars' emotions were inappropriate? \\
\textbf{Other} 
& Feedback: Environment & Positive or negative aspects of the virtual environment. \\
& Feedback: Missing & Are there any features you felt were missing? \\
& Feedback: Suggestions & Do you have any feedback or suggestions for improvement? \\
& Recommend to Friend & Would you recommend this app to a friend who needed it? \\
\bottomrule
\multicolumn{3}{l}{\footnotesize $^*$ Note: In the analysis, this item was reverse-scored as \enquote{Technical Reliability}.}
\end{tabularx}
\end{table}

% \section{Results Visualisation}
% \label{appendix:viz}
% \begin{figure}[H]
% \centering
% \includegraphics[width = 0.9\hsize]{figures/Box_Avatar.png}
% \caption{Distribution of participant ratings for the Avatar Interaction dimension. The box plots illustrate the variance in scores for visual embodiment metrics. The solid and dashed lines within each box represent the median and mean scores, respectively. Notably, while "Child Avatar Realism" shows the lowest median, "Therapist Avatar Display" and "Child Avatar Satisfaction" demonstrate high consistency and overall positive reception.}
% \label{fig:box_avatar}
% \end{figure}

% (Rest of appendix content)

\end{document}